\documentclass[10pt,twocolumn,journal]{IEEEtran}

\usepackage{lettrine}
\usepackage{amsmath}
\usepackage{cases}
\usepackage{bm}
\usepackage{txfonts}
\usepackage{amssymb}
\usepackage{cite}
\usepackage{graphicx}
\usepackage{psfrag}
\usepackage{url}
\usepackage{stfloats}
\usepackage{algorithm}
\usepackage{algorithmic}        
\usepackage{multirow}
\usepackage{color}
\usepackage{romannum}
\usepackage{lipsum}
\usepackage{stfloats}
\usepackage[utf8]{inputenc}
\usepackage[skip=2pt,font=scriptsize]{caption}
\usepackage[skip=2pt,font=scriptsize]{subcaption}
\usepackage{tabularx}
\usepackage{booktabs}

\ifCLASSINFOpdf

\else
\fi

\begin{document}
\pagenumbering{arabic}
\title{Optimizing Wireless Networks with Deep Unfolding: Comparative Study on Two Deep Unfolding Mechanisms}

\author{Abuzar B. M. Adam \IEEEauthorrefmark{1},~\IEEEmembership{Member,~IEEE}, Mohammed A. M. Elhassan, Elhadj Moustapha Diallo
\thanks{\IEEEauthorrefmark{1}Corresponding author: Abuzar B. M. Adam  (abuzar@cqupt.edu.cn)}
\thanks{A. B. M. Adam and E. M. Diallo are with School of Communications and Information Engineering, Chongqing University of Posts and Telecommunications, Chongqing,
P. R. China, 400065.}
\thanks{M. A. M. Elhassan is with School of Mathematics and Computer Science, Zhejiang Normal University (e-mail: Mohammedac29@zjnu.edu.cn).}
}

\maketitle

\begin{abstract}
In this work, we conduct a comparative study on two deep unfolding mechanisms to efficiently perform power control in the next generation wireless networks. The power control problem is formulated as energy efficiency over multiple interference links. The problem is nonconvex. We employ fractional programming transformation to design two solutions for the problem. The first solution is a numerical solution while the second solution is a closed-form solution. Based on the first solution, we design a semi-unfolding deep learning model where we combine the domain knowledge of the wireless communications and the recent advances in the data-driven deep learning. Moreover, on the highlights of the closed-form solution, fully deep unfolded deep learning model is designed in which we fully leveraged the expressive closed-form power control solution and deep learning advances. In the simulation results, we compare the performance of the proposed deep learning models and the iterative solutions in terms of accuracy and inference speed to show their suitability for the real-time application in next generation networks
\end{abstract}


\section{Introduction}

\lettrine[lines=2]{A}{lthough} the fifth generation (5G) will be deployed in many countries and an intensive research work has been contributed to enhance it, many researchers and developers have pointed their attention into the six generation (6G) \cite{bb1}. Visions for 6G have been laid down and an initiative has been proposed in the international telecommunication union (ITU) to establish the standards for the 2030 and beyond wireless technologies \cite{bb2}.

The deployment of 6G is expected to meet technical features including high data rate that has the peak rate of 20 Gbps for downlink and 10 Gbps for uplink. Quality of experience in which the user receives a minimum data rate at any given moment regardless of their location within the network. This can be measured in the cell edge to reflect the quality of the network design. Additionally, latency is expected to improve in 6G compared to that of 5G taking into consideration the complication due the large number of entities in the network. For energy efficiency (EE) and spectral efficiency, the power consumption is expected to decrease and EE can improve 100 times compared to 5G while the spectral efficiency is expected to be three times higher \cite{bb1,abm5}. In this context, EE in next-generation networks, including 5G and beyond (6G and future technologies), is a critical area of focus due to the exponentially increasing demand for data and the global emphasis on reducing energy consumption and carbon footprint \cite{abm1,abm2,abm3,abm4}.

To meet the above requirements, intelligent communications is envisioned as a key enabler for 6G networks \cite{bb3}. Hence, deep learning (DL) is applied in to improve wireless communication by solving different problems such as channel estimation \cite{bb4,bb5,bb6,bb7,bb8,bb9}, optimal resource allocation \cite{bb10,bb11, bb12}. Based on the mechanism of employing DL in solving wireless communications problems, the proposed methods can be categorized into main three categories: data-driven methods, model-driven methods, joint model-data-driven models. In data-driven methods, algorithms are designed to find the patterns and the connections which are hidden in the data.

Recently, model-driven or deep unfolding has been introduced as a key to enhance the performance of communication system. This mechanism combines both domain knowledge with learning ability of DL to overcome the problem of both traditional methods and DL-based methods (i.e., data-driven methods) \cite{bb43}. Although, this mechanism has been mainly developed signal recovery \cite{bb44,bb45}, some studies used it for precoding design \cite{bb46,bb47} and beamforming \cite{bb48,bb49,bb50,mod1}. However, very few studies have been dedicated for model-based resource allocation. For instance, the study in \cite{bb51} investigated unfolding of WMMSE using graph neural network. The idea was to perform energy-efficient power allocation. They have shown the advantages of the deep unfolding mechanism over the data-driven methods in terms of time and computation cost.

In this work, we investigate the employment of deep unfolding to design energy-efficient power control in multicell networks. We also consider modern neural networks architectures in addition to deep unfolding to design a hybrid model. Therefore, this study provides comparative aspects to conceive and incorporate the advantages of both mechanisms. The motivations of this work can be summarized in the following points:
\begin{itemize}
  \item In the literature, although there are several studies on deep unfolding in precoding and signal recovery, but there are few studies dedicated for power control. We introduce this work to contribute in improving the work in this field.
  \item Although the study in \cite{bb47} incorporated DNN to handle part of the problem, however there is a room to explore the advantages of both data-driven deep learning and model-driven to deep unfolding methods. Our goal is to incorporate the advantages of other DL models when designing deep unfolding based models.
  \item The emphasis on specific patterns in multivariate system is one of domain knowledge nonnegligible trait in communications systems. The impact of several parameters should be incorporated during the design of DL models which can enhance the prediction \cite{bb38}. In this work, we want to show the advantage of multivariate patterns prediction in improving the model accuracy.
\end{itemize}
Our contributions can be summarized as follows:
\begin{itemize}
  \item We investigate power control in multicell networks. The problem is formulated as EE maximization. The formulated problem is sum-of-ratios problem (SoRP). Obtaining the global optimal solution in the polynomial time for this problem is difficult. Therefore, we propose multidimensional fractional programming-based power control frameworks. The proposed frameworks convert the problem into a sequence of convex problems using auxiliary variables. To perform this conversion, the auxiliary variables are used to separate the rate function from the total power consumption and the signal from the interference. Based on the above procedure, we design two solutions to handle the problem.
  \item Despite that the proposed fractional programming solutions have the advantage of exploring most of the possible solutions in the solution space, However, they are time consuming and require several iterations to converge which make them unsuitable for real-time application. Hence, we resort to DL to handle the power control.
  \item First, we propose deep unfolding model based on the closed-from solution of power control i.e., we unfold the iterative solution into layers where we determine the learnable parameters from the unfolded solutions.
  \item We design another an end-to-end model on the highlights of the numerical solution and the deep unfolding while incorporating multiple attention sub-neural networks to compensate the loss. The attention blocks are used to perform multivariate prediction. Hence, we incorporate the advantages of modern DL architectures and data-driven mechanism.
  \item An ablation study is conducted to investigate the optimal design architectures for both models and the impact of each part of the proposed models.
  \item The simulation results have shown the efficiency of the proposed deep learning models in terms of accuracy and the inference speed.
\end{itemize}
The remainder of this paper is organized as follows: In Section II, we discuss the system model and problem formulation. In Section III, we introduce the fractional programming based energy-efficient power allocation where we investigate the numerical solution and closed-form solution. Section IV, investigates both model-driven designs energy efficient power control. In Section V, we present our simulation results and discussion. In Section VI, we give the conclusions and the future works.

\section{System Model and Problem Formulation}

We consider a multi-cell interference network includes $M$ base stations (BSs) serving $K$ users. $\mathcal{M}$ and $\mathcal{K}$ represent the set of BS and users and denoted as ${\mathcal{M}} = \left\{ {1,2,....,M} \right\}$ and  ${\mathcal{K}} = \left\{ {1,2,....,K} \right\}$. We assume that each BS is equipped with $N$ antenna while each user is equipped with single antenna.
The BS $m$ transmits a signal to its associated users. Then, the received signal at the user $k$ associated with the $m$ is given as
\begin{equation}\label{eqn1}
{y_{m,k}} = \sqrt {\rho_{m,k}{P_{max}}} {\bf {h}}_{m,k}^H{{\bf {x}}_m} + \sum\limits_{n \in {\cal M},n \ne m} {\sum\limits_{k' \in {\cal K}} {\sqrt {{\rho_{n,k'}}P_{max}} {\bf {h}}_{n,k}^H{{\bf {x}}_n}} }  + {{\bf {n}}_{m,k}},
\end{equation}
where $\rho_{m,k}$ with $0<\rho_{m,k}<1$ is the power allocation coefficient of the user $k$ associated with BS $m$, and $P_{max}$ represents the maximum transmit power of the BS $m$, respectively. ${\bf{x}}_m$ is the transmit signal at the BS $m$ where $E\left[ {{{\left\| {{{\bf{x}}_m}} \right\|}^2}} \right] = 1$. ${{\bf {h}}_{m,k}} \in {\mathbb{C}^{N \times 1}}$ is the channel coefficient between the BS $m$ the user $k$. The second term in \eqref{eqn1} represents the intercell interference. ${\bf{n}}_{m,k}$ is the additive white Gaussian noise (AWGN) at the user $k$  with variance $\sigma^2$. Hence, the signal-to-interference-plus-noise ratio (SINR) of the user $k$ is expressed as,
\begin{equation}\label{eqn2}
{{{\gamma _{m,k}} = \frac{{{{\left| {{\bf{h}}_{m,k}^H} \right|}^2}{\rho _{m,k}}{P_{max}}}}{{\sum\limits_{j \ne k} {{{\left| {{\bf{h}}_{m,k}^H} \right|}^2}{\rho _{m,j}}{P_{max}}}  + \sum\limits_{n \in M,n \ne m} {\sum\limits_{k' \in K} {{{\left| {{\bf{h}}_{n,k}^H} \right|}^2}{\rho _{n,k'}}{P_{max}}} }  + \sigma _{m,k}^2}},}}
\end{equation}

In presence of intercell interference, the order of SINR is considered for power allocation and successive interference cancelation (SIC) in both BS and user[]. without loss of generality, we can assume that ${\gamma _{m,1}} > {\gamma _{m,2}} > ... > {\gamma _{m,K}}$. Using SINR information, power allocation coefficients are computed. After SIC is carried out, SINR of the user $k$ associated with BS $m$ can be expressed as,
\begin{equation}\label{eqn3}
{\gamma _{m,k}} = \frac{{{{\left| {{\bf{h}}_{m,k}^H} \right|}^2}{\rho _{m,k}}{P_{max}}}}{{\sum\limits_{j = 1}^{k - 1} {{{\left| {{\bf{h}}_{m,k}^H} \right|}^2}{\rho _{m,j}}{P_{max}}}  + \sum\limits_{n \in M,n \ne m} {\sum\limits_{k' \in K} {{{\left| {{\bf{h}}_{n,k}^H} \right|}^2}{\rho _{n,k'}}{P_{max}}} }  + \sigma _{m,k}^2}},
\end{equation}
And the achievable rate of the user $k$ is given as
\begin{equation}\label{eqn4}
{R_{m,k}} = B{\log _2}\left( {1 + {\gamma _{m,k}}} \right)
\end{equation}
Then, EE of the user $k$ is given as follows
\begin{equation}\label{eqn5}
{\eta _{m,k}} = \frac{{{R_{m,k}}}}{{{\rho_{m,k}}P_{max} + {p_{k,c}}}}
\end{equation}
where $p_{k,c}$ is the circuit power consumption.

We aim at maximize the weighted sum energy efficiency (WSEE) of the network. Unlike the global energy efficiency (GEE) where some links and users are favored, WSEE gives the opportunity to do the balance by explicitly prioritizing some links via different weight assignment \cite{bb52}. The optimization problem can be formulated as,
\begin{subequations}\label{eqn6:main}
\begin{align}
& \underset{\forall \rho_{m,k}}{\text{maximize}} & & \sum\limits_{m = 1}^M {\sum\limits_{k = 1}^K {{\omega _{m,k}}{\eta _{m,k}}} }& & \tag{\ref{eqn6:main}}\\
&\text{subject to}&& \sum\limits_{m \in {\cal K}} {{\rho _{m,k}}}  \le 1,\label{eqn6:a}
\end{align}
\end{subequations}
where ${\omega _{m,k}}$ is the weight. The problem is nonconvex and sum of ratio problem (SoRP). Therefore, it is difficult to obtain the global solution with limited complexity \cite{bb52,bb53}. Nevertheless, we can recast it as a tractable convex problem as in the Section III. Toward the goal of improving the network performance, Section IV includes design of deep learning model to perform EE maximization in the real-time. This is achieved by projecting the traditional solution in Section III to a deep learning model.

\section{Fractional Programming-based Energy-Efficient Power Allocation}
In this section, we consider a low complexity algorithms to suboptimally solve problem \eqref{eqn6:main}. The proposed solutions is a quadratic transform based fractional programming. For similar problems as \eqref{eqn6:main}, some studies such as \cite{bb52,bb54} employed successive convex approximation (SCA) to obtain the lower bound.  But the result of the this approximation on \eqref{eqn6:main} will be sum of pseudo-concave functions which is not guaranteed to be pseudo-concave \cite{bb38,bb52}. To obtain a transformed function that fulfills the desired properties, we resort to the fractional programming method in \cite{bb53}. Let us consider the following problem
\begin{equation}
\begin{aligned}
& \underset{x}{\text{maximize}}
& & \sum\limits_{i = 1}^F {f\left( {\frac{{{g_i}\left( x \right)}}{{{h_i}\left( x \right)}}} \right)}\\
& \text{subject to }
& &x \in {\cal X},
\end{aligned}
\label{eqn7}\end{equation}
According to this method, the transformed function should satisfy the following technical conditions:
\begin{itemize}
  \item The transformed objective function $\tilde f\left( {x,\lambda } \right)$ can be decoupled as $\tilde f\left( {x,\lambda } \right) = {f_1}\left( \lambda  \right)A\left( {{g_i}\left( x \right)} \right) + B\left( {{h_i}\left( x \right)} \right){f_2}\left( \lambda  \right)$ where $\lambda$ is an auxiliary variable.
  \item The function $\tilde f\left( {x,\lambda } \right)$ is concave over $\lambda$  when $x$ is fixed.
  \item $x^*$ maximizes $f\left(x\right)$ if and only if $x^*$ with some $\lambda^*$ maximize $\tilde f\left( {x,\lambda } \right)$.
  \item For ${\lambda ^*} = \mathop {\arg \max }\limits_y \tilde f\left( {x,\lambda } \right)$ given some $x$, then $\tilde f\left( {x,{\lambda ^*}} \right) = f\left( x \right)$ for this $x$.
\end{itemize}
\subsection{Numerical Solution for Power Allocation}
On the highlights of the above conditions, we are going to transform the objective function in \eqref{eqn6:main}. First applying quadratic transform, we obtain the following equivalent problem
\begin{subequations}\label{eqn8:main}
\begin{align}
& \underset{\forall \rho_{m,k}, y}{\text{maximize}} & & \sum\limits_{m = 1}^M {\sum\limits_{k = 1}^K {{\omega _{m,k}}\left( {2y{{\left( {{R_{m,k}}} \right)}^{\frac{1}{2}}} - {y^2}\left( {{\rho_{m,k}}P_{max} + {p_{k,c}}} \right)} \right)} }& & \tag{\ref{eqn8:main}}\\
&\text{subject to}&& \eqref{eqn6:a}\notag\\
&                 && {y_{m,k}} \in \mathbb{R},\label{eqn8:a}
\end{align}
\end{subequations}

where $y$ is the set of variables $\left\{ {{y_{1,1}},...,{y_{M,K}}} \right\}$. The optimal value of $y^*_{m,k}$ is given when $\rho_{m,k}$ is fixed as
\begin{equation}\label{eqn9}
y_{m,k}^* = \frac{{{{\left( {{R_{m,k}}} \right)}^{\frac{1}{2}}}}}{{{\rho_{m,k}}P_{max} + {p_{k,c}}}}
\end{equation}
Problem \eqref{eqn8:main} is equivalent to \eqref{eqn6:main} and first technical condition is met. The term ${\left( {{R_{m,k}}} \right)^{\frac{1}{2}}}$ is a non-decreasing and its concavity can be established if $R_{m,k}$ is linear or concave while $y_{m,k}$ can be seen as a maximizer point. Obviously, when $\rho_{m,k}$ is fixed, the objective function is concave over $y$. However, due to the interference,  $R_{m,k}$ is nonconcave. We employ multidimensional quadratic transform to linearize  $R_{m,k}$. This is done by separating the signal from the interference. Thus, we have the transformed function in \eqref{eqn10} on top of next page.
\begin{figure*}
\begin{equation}\label{eqn10}
{{\tilde R}_{m,k}}\left( {\rho ,z} \right) = B{\log _2}\left( {1 + 2{z_{m,k}}{{\left( {{{\left| {{\bf{h}}_{m,k}^H} \right|}^2}{\rho _{m,k}}{P_{max}}} \right)}^{\frac{1}{2}}} - {z_{m,k}^2}\left( {\sum\limits_{j = 1}^{k - 1} {{{\left| {{\bf{h}}_{m,k}^H} \right|}^2}{\rho _{m,j}}{P_{max}}}  + \sum\limits_{n \in M,n \ne m} {\sum\limits_{k' \in K} {{{\left| {{\bf{h}}_{n,k}^H} \right|}^2}{\rho _{n,k'}}{P_{max}}} }  + \sigma _{m,k}^2} \right)} \right),
\end{equation}
\end{figure*}

$z$ represents the set $\left\{ {{z_{1,1}},....,{z_{M,K}}} \right\}$. Where the optimal $z_{m,k}^*$ is given when other variables are fixed as follows
\begin{equation}\label{eqn11}
z_{m,k}^* = \frac{{{{\left( {{{\left| {{\bf{h}}_{m,k}^H} \right|}^2}{\rho _{m,k}}{P_{max}}} \right)}^{\frac{1}{2}}}}}{{\sum\limits_{j = 1}^{k - 1} {{{\left| {{\bf{h}}_{m,k}^H} \right|}^2}{\rho _{m,j}}{P_{max}}}  + \sum\limits_{n \in M,n \ne m} {\sum\limits_{k' \in K} {{{\left| {{\bf{h}}_{n,k}^H} \right|}^2}{\rho _{n,k'}}{P_{max}}} }  + \sigma _{m,k}^2}}
\end{equation}
The objective function is concave over $z_{m,k}$ when the variables $\rho_{m,k}$ and $y_{m,k}$ are fixed. Hence, the equivalent problem can be written as,
\begin{subequations}\label{eqn12:main}
\begin{align}
&\underset{\rho,y,z}{\text{maximize}} & & \sum\limits_{m = 1}^M {\sum\limits_{k = 1}^K {\tilde f^{FP}_{m,k}} }& & \tag{\ref{eqn12:main}}\\
&\text{subject to}&& \eqref{eqn6:a}, \eqref{eqn8:a}\notag\\
&                 && {z_{m,k}} \in \mathbb{R},\label{eqn12:a}
\end{align}
\end{subequations}

where $\tilde f^{FP}_{m,k}$ is given as follows
\begin{equation}\label{eqnf}
\tilde f^{FP}_{m,k}={\omega _{m,k}}\left( {2y_{m,k}{{\left( {{{\tilde R}_{m,k}}\left( {\rho,z} \right)} \right)}^{\frac{1}{2}}} - {y^2_{m,k}}\left( {{\rho_{m,k}}P_{max} + {p_{k,c}}} \right)} \right)
\end{equation}
The problem in \eqref{eqn12:main} is convex on $\rho_{m,k}$ when $z_{m,k}$ and $y_{m,k}$ are fixed and it can be solved using CVX \cite{bb55}. Algorithm 1 illustrates the logical procedure to solve \eqref{eqn12:main}.

\begin{algorithm}[H]
\caption {Proposed Iterative Numerical Solution for Problem \eqref{eqn12}}
\begin{algorithmic}[1]
\renewcommand{\algorithmicrequire}{\textbf{Initialization:}}
\REQUIRE $\rho_{m,k}$, $\epsilon$, $t=0$.
\WHILE{$|\tilde f^{FP}_{m,k}\left(t\right) - \tilde f^{FP}_{m,k}\left(t-1\right)|\ge\epsilon$}
\STATE Update $z_{m,k}$ according to \eqref{eqn11}.
\STATE Update $y_{m,k}$ according to \eqref{eqn9}.
\STATE Update $\rho_{m,k}$ by solving \eqref{eqn12:main} with fixed $z_{m,k}$ and $y_{m,k}$.
\STATE $t=t+1$.
\ENDWHILE
\end{algorithmic}
\end{algorithm}

\subsection{Closed-form Solution for Power Allocation}
Obtaining a closed-form solution for the power allocation is desirable to design a deep unfolding model. However, the solution of power allocation problem \eqref{eqn12:main} can not be expressed in a closed-form. To circumvent this issue, implement the following steps.

First, we apply Lagrange dual transform \cite{bb56} on the data rate function, we get \eqref{eqn13} on top of next page.
\begin{figure*}
\begin{equation}\label{eqn13}
{R_{m,k}}\left( {\rho ,\gamma } \right) = B{\log _2}\left( {1 + {\gamma _{m,k}}} \right) - B{\gamma _{m,k}} + \frac{{B{{\left| {{\bf{h}}_{m,k}^H} \right|}^2}{\rho _{m,k}}{P_{max}}\left( {1 + {\gamma _{m,k}}} \right)}}{{\sum\limits_{j = 1}^{k - 1} {{{\left| {{\bf{h}}_{m,k}^H} \right|}^2}{\rho _{m,j}}{P_{max}}}  + \sum\limits_{n \in M,n \ne m} {\sum\limits_{k' \in K} {{{\left| {{\bf{h}}_{n,k}^H} \right|}^2}{\rho _{n,k'}}{P_{max}}} }  + \sigma _{m,k}^2}},
\end{equation}
\end{figure*}

Next, applying multidimensional quadratic transform, ${R_{m,k}}\left( {\rho,\gamma } \right)$ can be written as
\begin{equation}\label{eqn14}
\begin{aligned}
{R_{m,k}}\left( {\rho,\gamma ,z} \right) =& B{\log _2}\left( {1 + {\gamma _{m,k}}} \right) - B{\gamma _{m,k}} - z_{m,k}^2\\
&+ 2z_{m,k}\sqrt {{B{{\left| {{\bf{h}}_{m,k}^H} \right|}^2}{\rho _{m,k}}{P_{max}}\left( {1 + {\gamma _{m,k}}} \right)}} \\
& - z_{m,k}^2\left( \begin{array}{l}
\sum\limits_{j = 1}^{k - 1} {{{\left| {{\bf{h}}_{m,k}^H} \right|}^2}{\rho _{m,j}}{P_{max}}} \\
 + \sum\limits_{n \in M,n \ne m} {\sum\limits_{k' \in K} {{{\left| {{\bf{h}}_{n,k}^H} \right|}^2}{\rho _{n,k'}}{P_{max}}} }  + \sigma _{m,k}^2
\end{array} \right),
\end{aligned}
\end{equation}
where $z_{m,k}^*$ is given as follows
\begin{equation}\label{eqn15}
z_{m,k}^* = \frac{{\sqrt {{B{{\left| {{\bf{h}}_{m,k}^H} \right|}^2}{\rho _{m,k}}{P_{max}}\left( {1 + {\gamma _{m,k}}} \right)}} }}{\sum\limits_{j = 1}^{k - 1} {{{\left| {{\bf{h}}_{m,k}^H} \right|}^2}{\rho _{m,j}}{P_{max}}}  + \sum\limits_{n \in M,n \ne m} {\sum\limits_{k' \in K} {{{\left| {{\bf{h}}_{n,k}^H} \right|}^2}{\rho _{n,k'}}{P_{max}}} }  + \sigma _{m,k}^2},
\end{equation}
Substituting \eqref{eqn14} in \eqref{eqn8:main}, the final optimization problem is given as follows
\begin{subequations}\label{eqn16:main}
\begin{align}
& \underset{\rho,\gamma,z,y}{\text{maximize}} & & \sum\limits_{m = 1}^M {\sum\limits_{k = 1}^K {f^{CF}_{m,k}} }& & \tag{\ref{eqn16:main}}\\
&\text{subject to}&& \eqref{eqn6:a},\notag\\
&                 && {y_{m,k}},{z_{m,k}} \in \mathbb{R}, \label{eqn16:a}
\end{align}
\end{subequations}
where $f^{CF}_{m,k}$ is given as follows
\begin{equation}\label{eqnf2}
f^{CF}_{m,k}= \left( {2{y_{m,k}}{R_{m,k}}\left( {\rho,\gamma ,z} \right) - y_{m,k}^2\left( {{\rho_{m,k}}P_{max} - {p_{k,c}}} \right)} \right)
\end{equation}
The problem \eqref{eqn16:main} is convex in $p_{m,k}$ when other variables are fixed. The solution for $p_{m,k}$ can be expressed in a closed-form by taking the derivative for the objective function in \eqref{eqn16:main} and set it to zero. Hence, the closed-form solution is given as in \eqref{eqn17} on to of next page, where $I_{m,k}$ is given as
\begin{figure*}
\begin{equation}\label{eqn17}
{\rho _{m,k}} = \min \left\{ {\frac{{{{\left[ {{{\left( {{z_{m,k}}\left( {1 + {\gamma _{m,k}}} \right){{\left| {{\bf{h}}_{m,k}^H} \right|}^2}{P_{max}}} \right)}^2} - y_{m,k}^2\left( {{{\log }_2}\left( {1 + {\gamma _{m,k}}} \right) + {\gamma _{m,k}} + z_{m,k}^2\left( {1 - I_{m,k} - \sigma _{m,k}^2} \right)} \right)} \right]}^2}}}{{4y_{m,k}^4z_{m,k}^2\left( {1 + {\gamma _{m,k}}} \right){{\left| {{\bf{h}}_{m,k}^H} \right|}^2}{P_{max}}}},1} \right\},
\end{equation}
\end{figure*}

\begin{equation}\label{eqn18}
I_{m,k} = \sum\limits_{j = 1}^{k - 1} {{{\left| {{\bf{h}}_{m,k}^H} \right|}^2}{\rho _{m,j}}{P_{max}}}  + \sum\limits_{n \in M,n \ne m} {\sum\limits_{k' \in K} {{{\left| {{\bf{h}}_{n,k}^H} \right|}^2}{\rho _{n,k'}}{P_{max}}} },
\end{equation}

\begin{algorithm}[H]
\caption {Proposed Iterative Solution for Problem \eqref{eqn16:main}}
\begin{algorithmic}[1]
\renewcommand{\algorithmicrequire}{\textbf{Initialization:}}
\REQUIRE $\rho_{m,k}$, $\epsilon$, $t=0$.
\WHILE{$|f^{CF}_{m,k}\left(t\right) - f^{CF}_{m,k}\left(t-1\right)|\ge\epsilon$}
\STATE Update $z_{m,k}$ according to \eqref{eqn15}.
\STATE Update $\gamma_{m,k}$ using \eqref{eqn3}.
\STATE Update $y_{m,k}$ according to \eqref{eqn9}.
\STATE Update $\rho_{m,k}$ by solving \eqref{eqn17} with fixed $z_{m,k}$, $\gamma_{m,k}$, and $y_{m,k}$.
\STATE $t=t+1$.
\ENDWHILE
\end{algorithmic}
\end{algorithm}

\emph{Theorem 1:} Algorithm 1 and Algorithm 2 converge to stationary point of \eqref{eqn6:main}.

\emph{Proof:} Please, refer to Appendix A.

Obtaining the solution for \eqref{eqn12:main} requires solving the problem and updating the parameters for all users, this procedure requires
$O\left( {K\left( {3M + 1} \right)\log \left( {{\raise0.7ex\hbox{$1$} \!\mathord{\left/
 {\vphantom {1 }}\right.\kern-\nulldelimiterspace}
\!\lower0.7ex\hbox{$$}}} \epsilon\right)} \right)$. Algorithm 2 performs the update for one additional step. Hence, Algorithm 2 entails an asymptotic complexity $O\left( {K\left( {3M + 2} \right)\log \left( {{\raise0.7ex\hbox{$1$} \!\mathord{\left/
 {\vphantom {1 }}\right.\kern-\nulldelimiterspace}
\!\lower0.7ex\hbox{$$}}} \epsilon\right)} \right)$ in the worst case.

\section{Model-based Design for Energy-Efficient Power Control}
In this section, we discuss the structure of the proposed model-based design techniques for the energy-efficient power control. Two models are designed; the first model is a semi-unfolding based model in which we unfold Algorithm 1 while in the second model we fully unroll Algorithm 2 to design the model.

\begin{figure}[!ht]
\centerline{\includegraphics[width=\columnwidth]{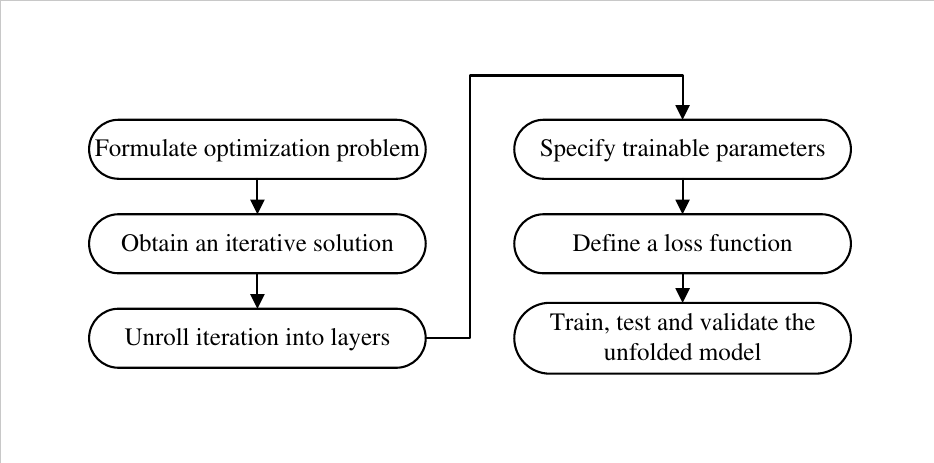}}
\caption{A typical deep unfolding procedure.}
\label{fig1}
\end{figure}

The main idea of deep unfolding is the unrolling of the iterations of principled inference algorithm to form a layered neural network \cite{bb43}. Hence, in deep unfolding-based power control, we employ the main domain knowledge in wireless resource allocation and deep leaning to build more efficient and low complexity models. Fig. \ref{fig1} shows the steps of the typical deep unfolding process.

The existence of the closed-form solution for the power allocation helps in designing a full unfolding-based model. However, this is difficult to apply in case of Algorithm 1 where the final form of the problem is solved numerically.
\subsection{Semi-Unfolding based Energy-Efficient Power Control}

\begin{figure*}
  \includegraphics[width=\textwidth,height=8cm]{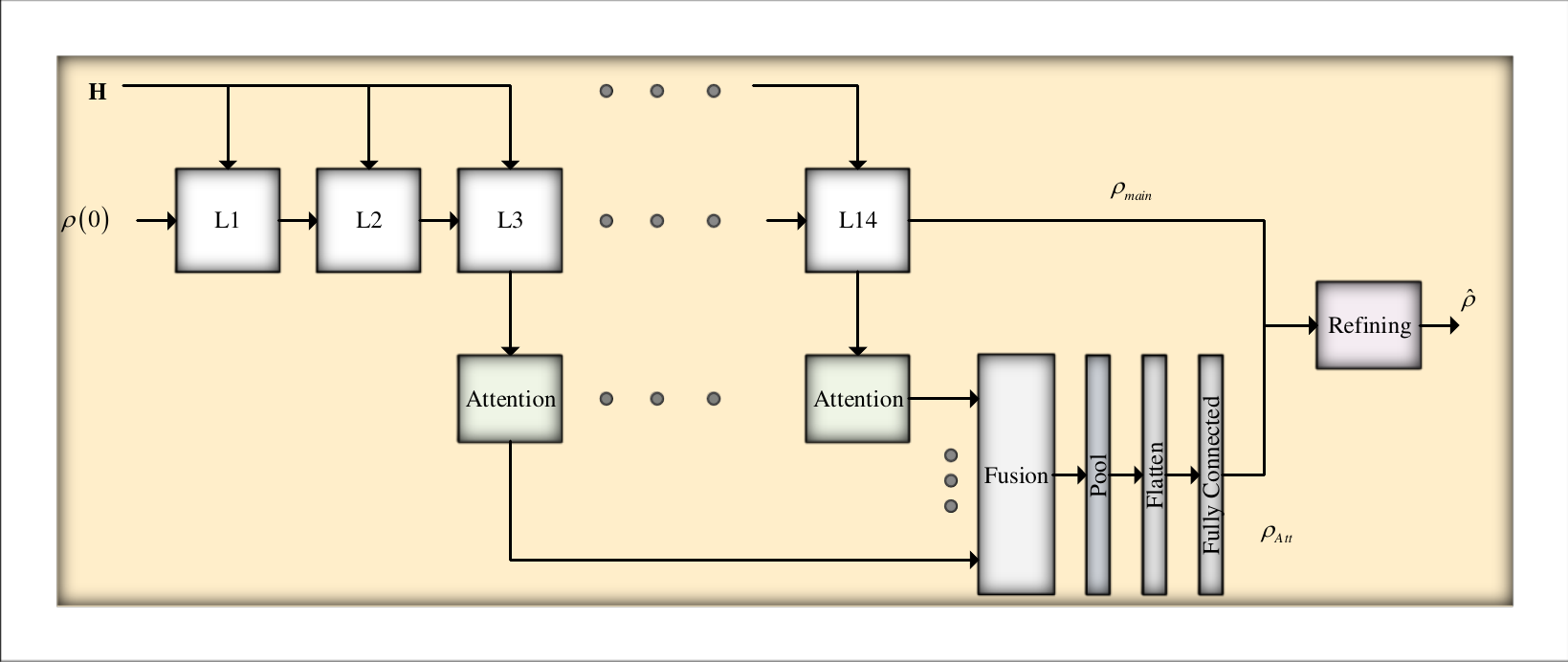}
  \caption{Structure of the proposed model.}
  \label{ovll}
\end{figure*}
Fig. \ref{ovll} shows the overall structure of the proposed fractional programming based multi-attention-based semi-unfolding model (MASUM). When projecting the above definition and concepts on our case, we can conceptualize the model design as follows
\begin{itemize}
  \item Domain knowledge embedded design: Algorithm 1 requires initialization of power and channel gain as input. The proposed algorithm has the advantage of separating the signal and the interference and converting the original problem into subproblems using the auxiliaries $y$ and $z$. The channel coefficients vector is considered as the main input to MASUM in addition to the initial power. Since $y$ and $z$ are functions of $p$, sub-networks are designed to represent their corresponding equations \eqref{eqn9} and \eqref{eqn11} as in the illustration of MASUM stages in Fig. \ref{stages}.
\begin{figure}[!ht]
\centerline{\includegraphics[width=\columnwidth]{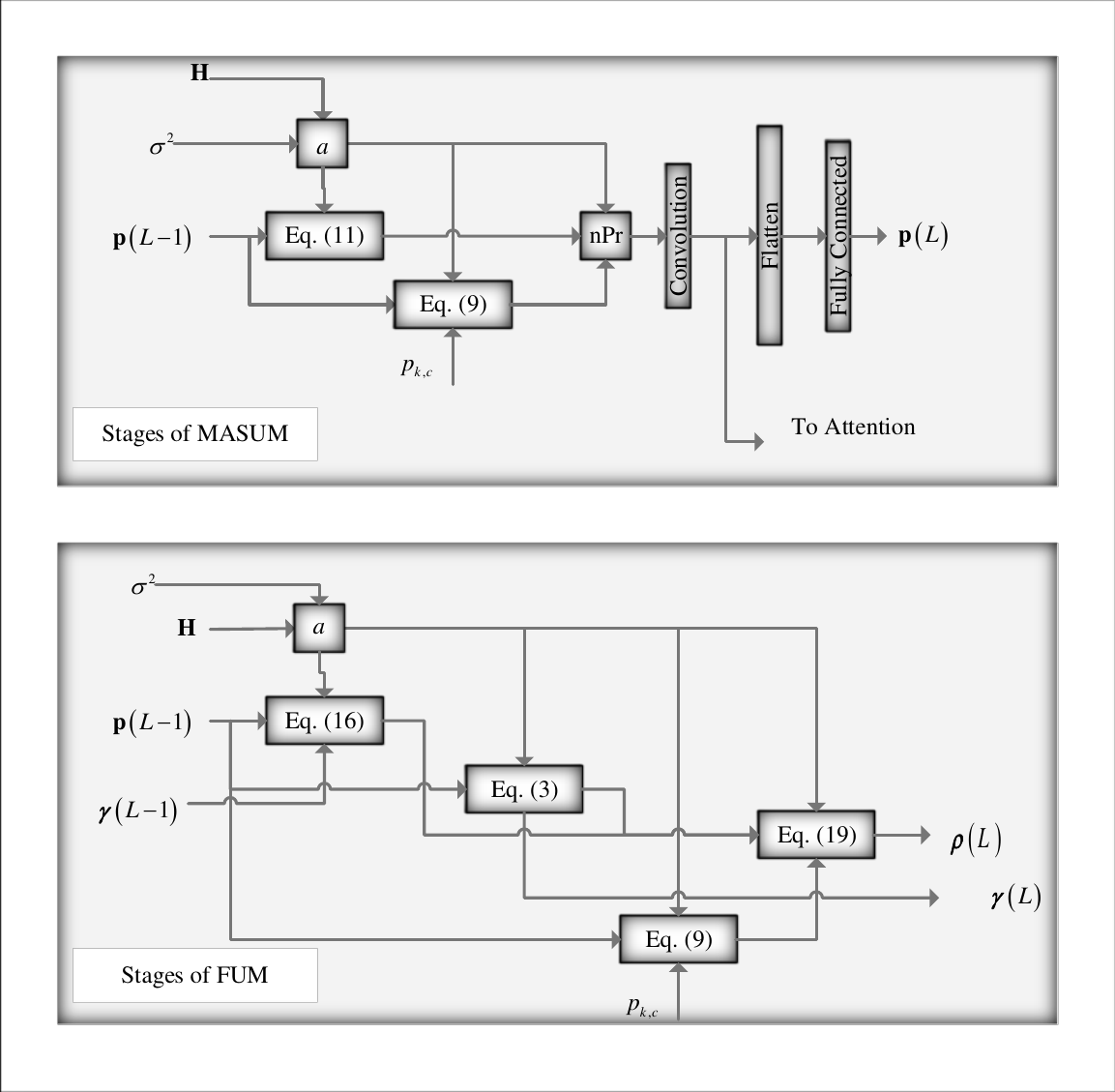}}
\caption{Detailed structure of stages of the proposed two models.}
\label{stages}
\end{figure}
  \item Leveraging deep learning advances: Due to the difficulties of fully representing the solution of \eqref{eqn12:main} using the generic deep unfolding mechanism, we resort to the data-driven advances to overcome this problem. As in Fig. \ref{stages}, convolutional layer is designed to create feature map for prediction. To increase the accuracy of the prediction, we design a permutation subnet \cite{bb57} to augment the data and enrich the input with more possible variations. The convolutional feature map is cloned and fed parallelly into two pipelines. The first pipeline includes flattening layer to reshape the features and a fully connected layer to convey power prediction.

      In the second pipeline, multiple attention blocks take the convolutional features and create feature maps where the impact of each single parameter is coded and the patterns are learnt and spatially connected \cite{bb38}.

      Given the convolutional feature map $f_c$, we apply parallel three $1\times1$ Conv layers on the $f_c$. Thus, we obtain the outputs $f_c^1$, $f_c^2$, and $f_c^3$. $f_c^1$ and $f_c^2$ are multiplied and inserted into a softmax as follows
       \begin{equation}\label{eqn20}
       f_s=softmax\left(f_c^1\cdot f_c^2\right)
       \end{equation}
       $f_s$ is the interaction between the different positions in the data where each element in $f_s$ is a weighted sum of features of other elements. Then, $f_s$ is used to calibrate the impact of each element in the feature map $f_c$ as follows
       \begin{equation}\label{eqn21}
       f_c^{cal}=f_s\cdot f_c^3
      \end{equation}
      $f_c^{cal}$ is reshaped and $1\times1$ Conv is applied where $\hat f_c^{cal}$ is obtained. The final output feature map of the attention block is as
      \begin{equation}\label{eqn22}
      f_{A}=\hat f_c^{cal} + f_c
     \end{equation}
       Multiple attention feature maps are fused using fusion network. To perform feature fusion, multiple attention feature maps are concatenated and fed into two $1\times1$ Conv layers to deliver the fusion map $f^{fusion}$.

       The fused features are fed into pooling and flattening layer for reshaping. Then, a fully connected layer is used to predict the power $\textbf{p}_{Att}$. The second pipeline can be seen as compensation of the prediction loss in the first pipeline.

      The power predictions from the two pipelines are refined to convey the final power prediction $\hat{\textbf{p}}$. Where $\hat{\textbf{p}}$ is given as
      \begin{equation}\label{eqn23}
      \hat{\textbf{p}} = \sigma\left(\textbf{p}_{Att} + \textbf{p}_{main}\right)\cdot \left(\textbf{p}_{Att} + \textbf{p}_{main}\right)
     \end{equation}

      Fig. \ref{eblocks} shows the detailed structure of the attention block, fusion block, and refining block.
     \begin{figure}[!ht]
      \centerline{\includegraphics[width=\columnwidth]{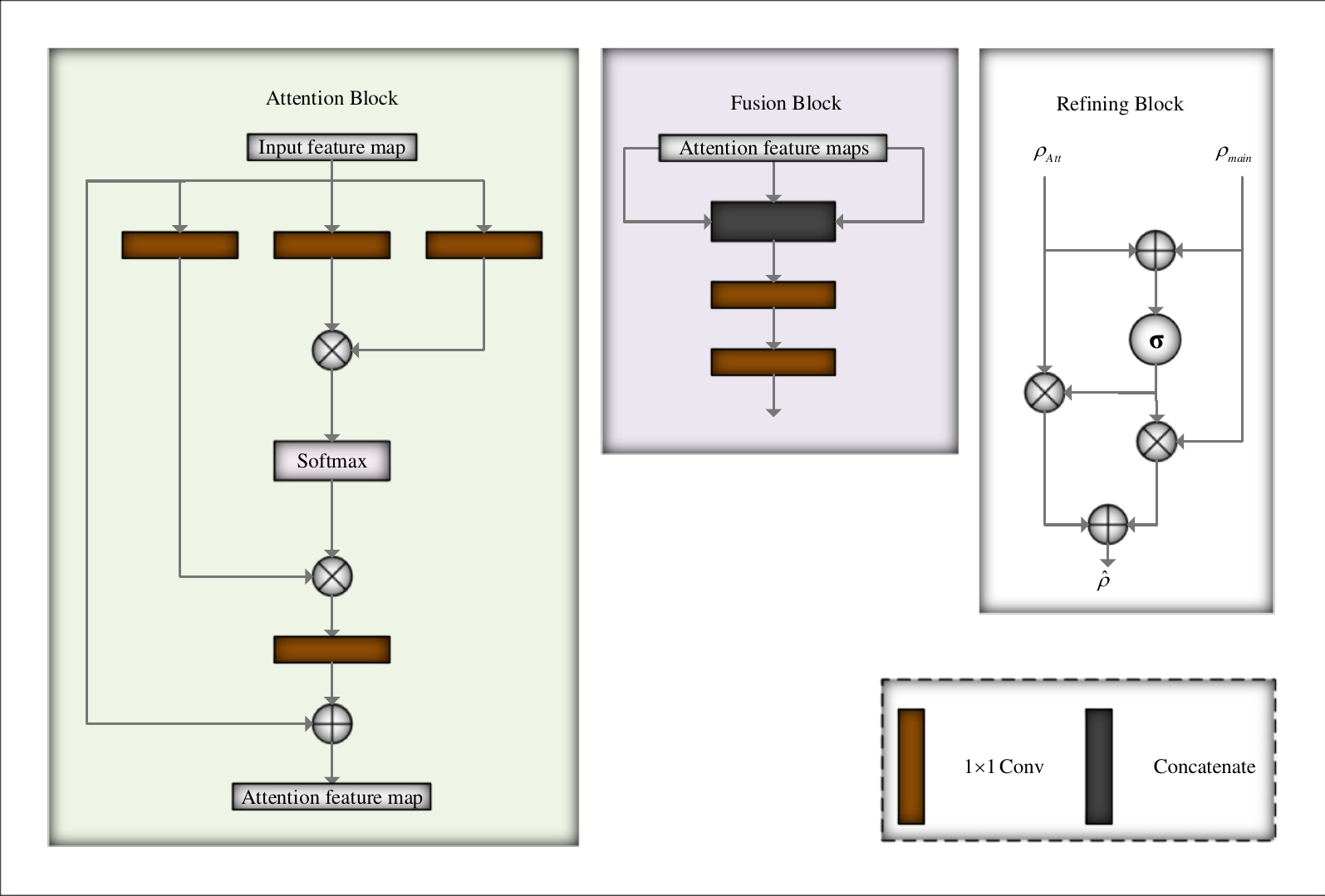}}
      \caption{Detailed building blocks for attention, fusion, and refining blocks.}
      \label{eblocks}
      \end{figure}
\end{itemize}

\subsection{Full-Unfolding based Energy-Efficient Power Control}
\begin{figure}[!ht]
\centerline{\includegraphics[width=\columnwidth]{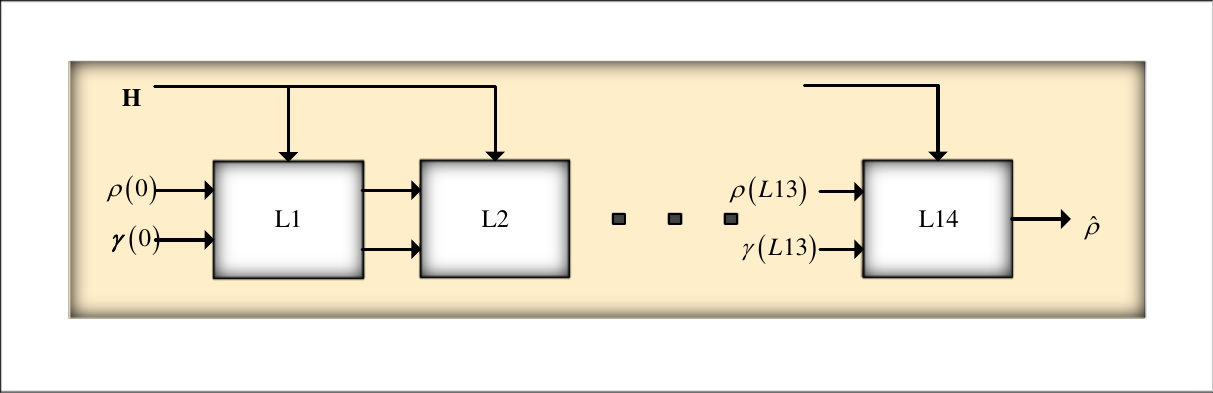}}
\caption{Structure of the proposed FUM.}
\label{full}
\end{figure}

The closed-form solution in \eqref{eqn17} is helpful in fully unfolding Algorithm 2. Fig. \ref{full} illustrates the block diagram of the full-unfolding based model(FUM). In Fig. \ref{stages}, the second block represents the structural illustration of the single stage of FUM.

The structure in Fig. \ref{full} can be viewed as DNN with number of layers equals to the iteration of Algorithm 2. The dynamics of each layer are described as in the second block in Fig. \ref{stages}. The structure can be viewed as a graph with shared variables among the computation nodes. To optimize the parameters, we employ the well-known technique of backpropagation. To efficiently optimize the model in Fig. \ref{full}, employ the available data and the underlaying distribution of the channel realization of in $\bf {G}$.
\subsection{Technical Description and Inference Process}
In this subsection, we give the technical description of the proposed two models
\subsubsection{MASUM Case}
\begin{itemize}
  \item \emph{Forward Propagation:} Provided with the channel coefficients the initial power, each block in the main pipeline in Fig. \ref{ovll} outputs a power vector. However, due to the presence of the data-driven layers (namely, the convolution, flatten, and fully connected layers), the model exhibits some performance loss. Hence, an effective attention-based loss compensation mechanism in the second pipeline is adopted. The forward propagation ${\cal F}\left(  \cdot  \right)$ of MASUM can be expressed as
      \begin{equation}\label{eqn24}
       {{\hat{\textbf{p}}}} = {\cal F}\left( {\left\{ G \right\};{\textbf{p}}} \right)
      \end{equation}
  \item \emph{Backpropagation:} In this procedure, the gradients of the learnable parameters are computed following the derivative of the loss function ${\cal L}\left(  \cdot  \right)$ \cite{bb58}.
  \item \emph{Updating Parameters:} Since each block in Fig. \ref{ovll} receives power input from the previous block, the update of the parameters is achieved based on \eqref{eqn9} and \eqref{eqn11} in addition to the gradient over ${\cal L}\left(  \cdot  \right)$ with respect to the power.
  \item \emph{Loss Function Characterization:} The loss function can be expressed as
  \begin{equation}\label{eqn25}
  {\cal L}\left( {\left\{ {G,{\textbf{p}}} \right\};{{\hat{\textbf{p}}}}} \right) \buildrel \Delta \over = \tilde f\left( {G,{\cal F}\left( {\left\{ G \right\};{\textbf{p}}} \right);{{\hat{\textbf{p}}}}} \right)
  \end{equation}
  where ${\tilde f}\left(  \cdot  \right)$ given in \eqref{eqnf}.
\end{itemize}
During the inference process, in each layer, the loss function is optimized in such a way that the power prediction is improved. However, performance loss is inevitable due to the data-driven. This loss is compensated by employing the changes throughout the main pipeline via the attention pipeline.
\subsubsection{FUM Case}
The learnable parameters in this case are $\textbf{p}$ and $\boldsymbol{\gamma}$. Each block outputs the $\textbf{p}$ and $\boldsymbol{\gamma}$  while the final block conveys only  $\textbf{p}$. The forward propagation ${\cal F}\left(  \cdot  \right)$ in this case can be expressed as
\begin{equation}\label{eqn26}
\left\{ {{{\hat{\textbf{p}}}},\boldsymbol{\gamma} } \right\} = {\cal F}\left( {G;\left\{ {{\textbf{p}},\boldsymbol{\gamma} } \right\}} \right)
\end{equation}
In the backpropagation, the gradients of the loss function with respect to the learnable parameters are derived using the chain rule. The loss function ${\cal L}\left(  \cdot  \right)$  can be expressed as
\begin{equation}\label{eqn27}
{\cal L}\left( {G,\left\{ {{\textbf{p}},\boldsymbol{\gamma} } \right\};{{\hat{\textbf{p}}}}} \right) \buildrel \Delta \over = f\left( {G,{\cal F}\left( {G;\left\{ {{\textbf{p}},\boldsymbol{\gamma} } \right\}} \right);{{\hat{\textbf{p}}}}} \right),
\end{equation}
where ${f}\left(  \cdot  \right)$ is defined in \eqref{eqnf2}. The updates of the learnable parameters at the sequence $t$ can be expressed as
\begin{equation}\label{eqn28}
{\boldsymbol{\gamma} ^{\left( {t + 1} \right)}}\left( L \right) = {\boldsymbol{\gamma} ^{\left( t \right)}}\left( L \right) - \delta \frac{{\partial {\cal L}\left( {G,\left\{ {{{\textbf{p}}},\boldsymbol{\gamma} } \right\};{{{{\hat{\textbf{p}}}}}^{\left( t \right)}}} \right)}}{{\partial \boldsymbol{\gamma} }}
\end{equation}
\begin{equation}\label{eqn29}
{{\textbf{p}}^{\left( {t + 1} \right)}}\left( L \right) = {{\textbf{p}}^{\left( t \right)}}\left( L \right) - \delta \frac{{\partial {\cal L}\left( {G,\left\{ {{\textbf{p}},\boldsymbol{\gamma} } \right\};{{{{\hat{\textbf{p}}}}}^{\left( t \right)}}} \right)}}{{\partial {\textbf{p}}}}
\end{equation}
where $\delta$ is the learning rate. From Fig. \ref{stages}, the above update is also equivalent to \eqref{eqn3} and the first term in \eqref{eqn17}.

Different from MASUM, during the inference, FUM optimizing the loss function straightforwardly while leveraging the full domain knowledge of the problem and the statistical information.
\subsection{Training and Testing Process}
Both of the models can be trained using the incremental learning procedure. In the increment round $\tau  \in \left\{ {0,...,{\rm T}} \right\}$, we optimize the cost function for the layer $l \in \left\{ {0,...,L14} \right\}$
by learning the learnable parameters of each model with mini-batches. At the end of increment round $\tau$, the layer $l=l+1$ is added to the network and a new round of training starts. The training results of learnable parameters of the layer $l$ are considered as initial values for the next round. At the final round, the entire model goes through a full training round. This training mechanism is proven to be the most effective training procedure for deep unfolding-based model.

For the simulation, we generate 22000 samples where 8000 samples are used for training where we run each of Algorithm 1 and Algorithm 2 multiple times to retain the best results. We generate the channel coefficients using the complex Gaussian distribution ${\cal C}{\cal N}\left( {0,1} \right)$ and the number of antennas  $N = 8$.

The batch size is set to 128 and 64 for MASUM and SUM respectively. The learning rate is set to 0.001 for both models. For MASUM, the kernel size is 3 in the main stage while the stride is set to 1.  The simulations are performed with Python 3.6, PyTorch 2.0, CUDA 10.1, cuDNN V8 on a desktop Intel Core i7-8700 CPU @ 3.20GHz, 8.00 GB RAM, and NVIDIA GeForce RTX 3050 8GB.
\section{SIMULATION RESULTS}
\begin{table}
\centering
\caption{Simulation Parameters}
\label{sim}
\begin{tabular}{|l|l|}
  \hline
  \textbf{Parameter} & \textbf{Value} \\
  \hline
   Number of users per BS &	2,4,8\\
   \hline
   Number of BSs &	4,7\\
   \hline
   Number of antennas $N$ & 2\\
   \hline
   Fast-Fading &	$Z \sim \mathcal{CN}\left(0,1\right)$\\
   \hline
  Path loss exponent &	3.5\\
  \hline
   Number of BS antennas &	8\\
   \hline
  $B$ &	250 kHz\\
  \hline
  Carrier frequency	& 2.4 GHz\\
  \hline
  Circuit power &	20 dBm\\
  \hline
\end{tabular}
\end{table}

In this section, we introduce the simulation results and performance evaluation for the proposed models. Each cell radius is 500 m and the adopted channel model follow the in force ITU-R P.1411-11. The users are randomly and uniformly distributed in the cells. The subchannel frequency is 250 kHz and the carrier frequency is 2.4 GHz. The noise power spectral density is -174 dBm/Hz and the noise figure is 7dB. The simulation parameters are deemed default as shown in TABLE \ref{sim} unless stated otherwise.

For comparison, we considered the deep learning model in \cite{bb38} as a black-box model (pure data-driven model). The black-box model is a multi-modal deep convolutional self-attention neural network. The black-box model is trained with multivariate data generated from Algorithm 1.
\subsection{Accuracy and Convergence}
\begin{figure}[htp]
\centering
\includegraphics[width=3.5in]{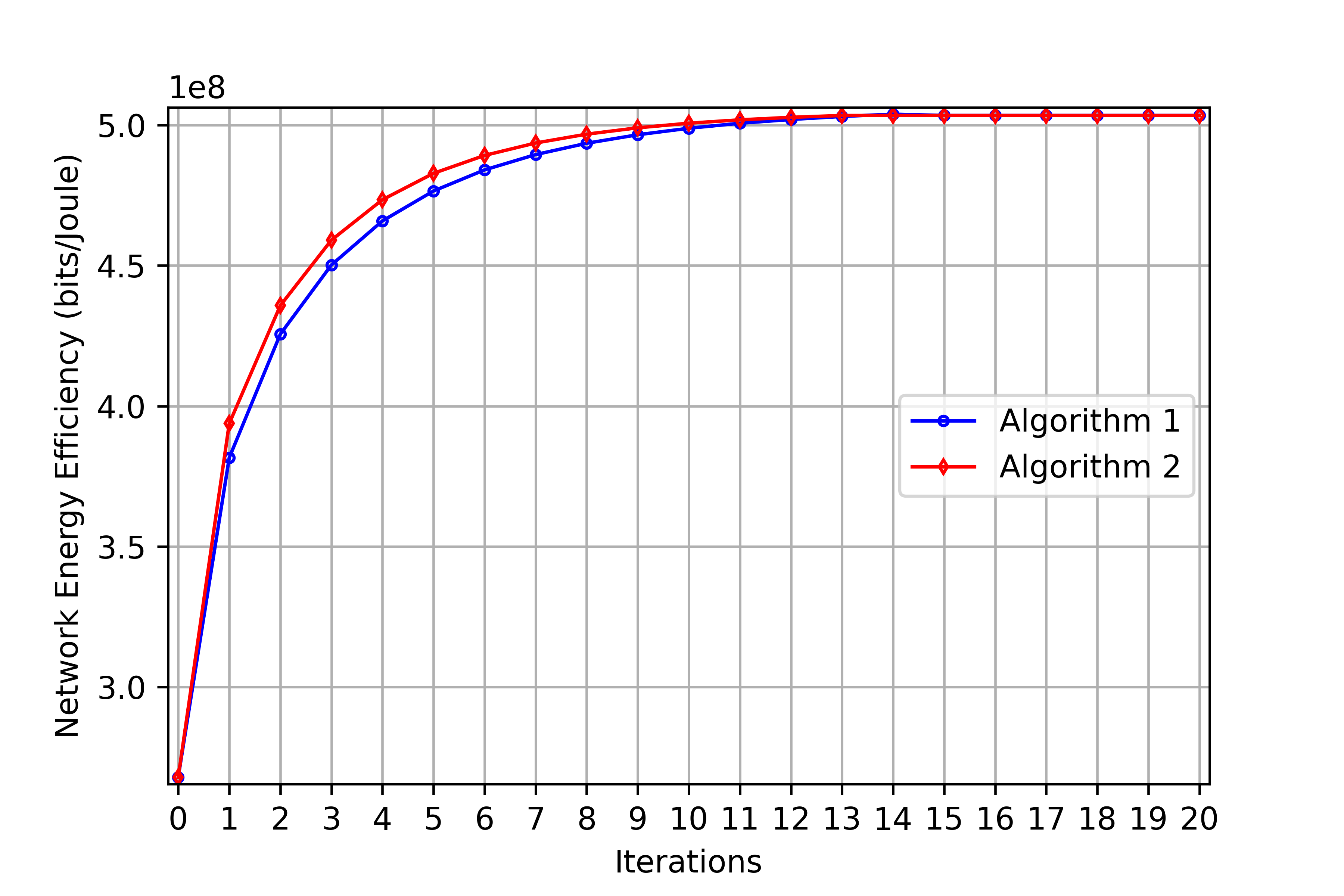}
\caption{Convergence of Algorithm 1 and Algorithm 2.}
\label{conv}
\end{figure}

From Theorem 1, both Algorithm 1 and Algorithm 2 converge to a fixed point of the original problem \eqref{eqn6:main}. Fig. \ref{conv} illustrates the convergence of both algorithms. The required number of iterations until convergence is 14 and 13 respectively for Algorithm 1 and Algorithm 2. Despite the slow convergence, both algorithms have the advantage of exploring the solution space.

From the complexity discussion of both algorithms, the number of users has the most impact on the running time of both algorithms. For instance, running both algorithms on CPU for 4 users per BS, we obtain the final output in 25.44 ms and 25.56 ms respectively for Algorithm 1 and Algorithm 2. In case of running both algorithms on GPU, we have the running time 10.04 ms and 10.01 ms respectively for Algorithm 1 and Algorithm 2.

\begin{figure}[htp]
\centering
\includegraphics[width=3.5in]{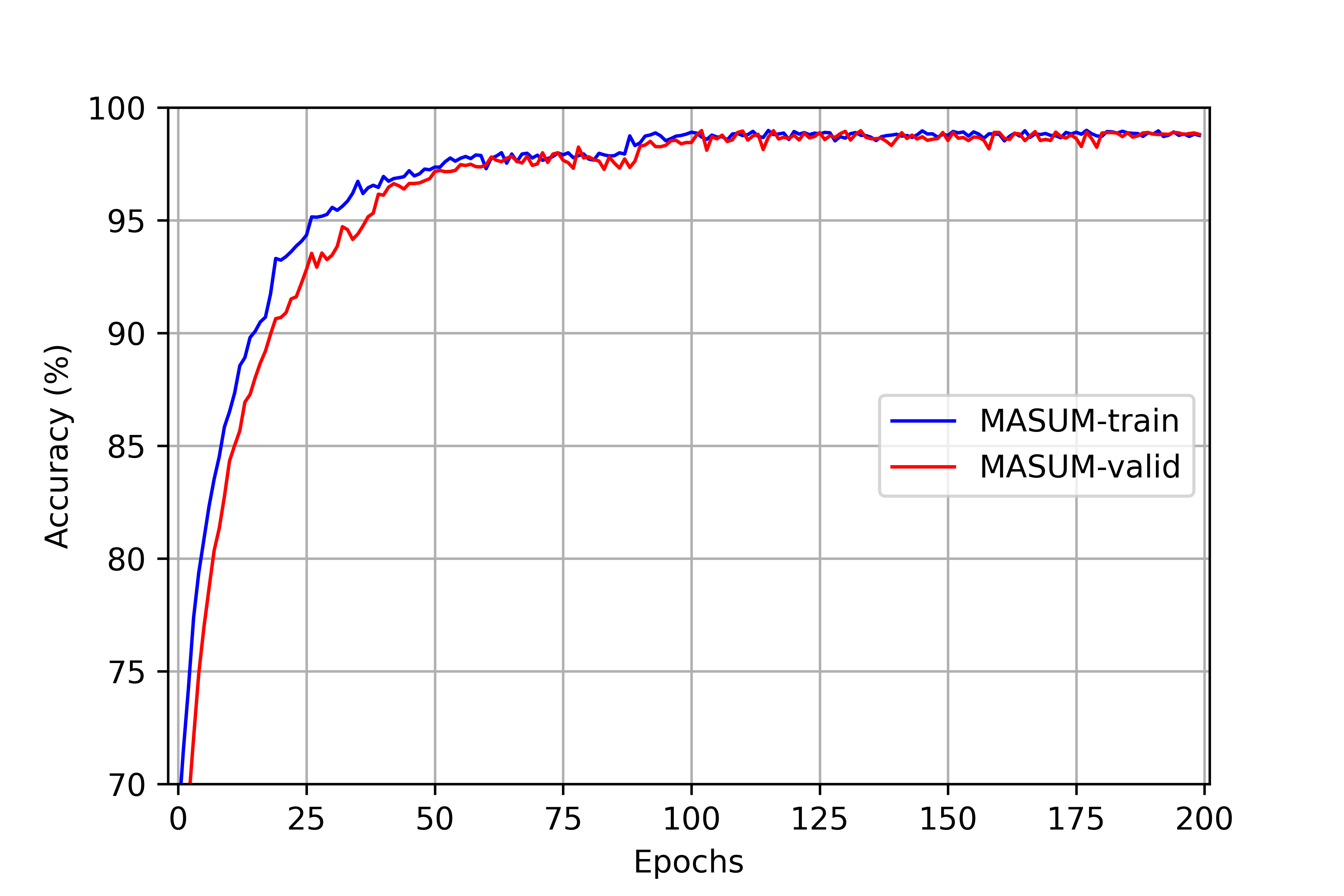}
\caption{Accuracy of MASUM.}
\label{accMASUM}
\end{figure}

Fig. \ref{accMASUM} shows the training and validation accuracy of MASUM. MASUM has the advantage of partially exploiting the domain knowledge of the problem and the available data. However, due to the presence of the data-driven parts in the model, the presence of the attention is advantageous as we can see later in the ablation study (see Section V-E). MASUM achieves training accuracy 98.80\% and validation accuracy of 98.78 for the case of full exploitation of model number of layers (e.g., 9 layers).

\begin{figure}[htp]
\centering
\includegraphics[width=3.5in]{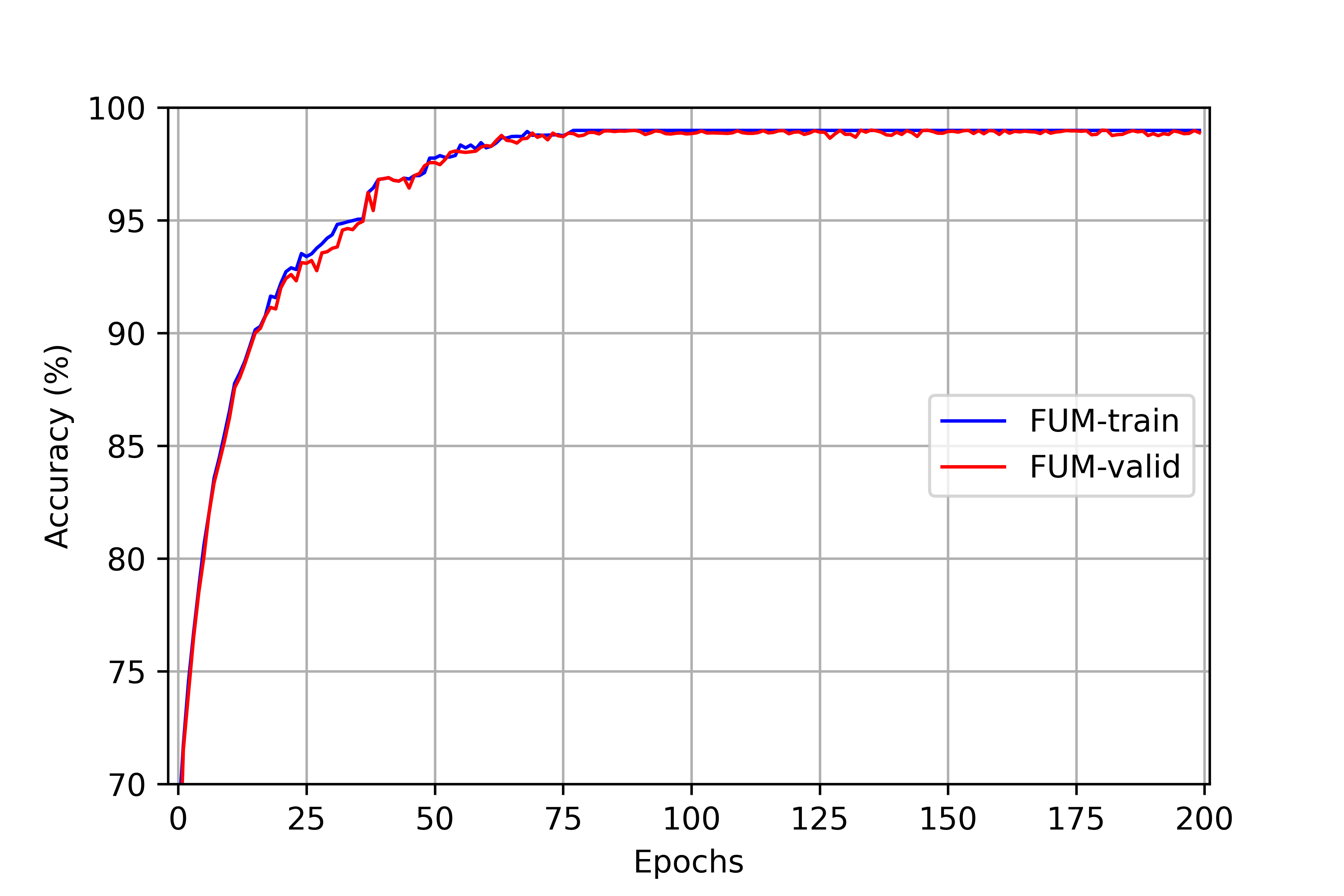}
\caption{Accuracy of FUM.}
\label{accFUM}
\end{figure}

Different from MASUM, FUM is fully exploiting both domain knowledge of the problem and the given available data. As in Fig. \ref{accFUM}, FUM achieves accuracy 99.32\% and 99.30\% respectively for training and validation. The proposed MASUM and FUM has a very quick inference since it does not require solving the problem as in the case Algorithm 1 and Algorithm 2. Hence, both MASUM and FUM exhibit great potential for real time applications. More details on the inference speed are revealed in Section V-E.

\subsection{Power Budget}
\begin{figure}[htp]
\centering
\includegraphics[width=3.5in]{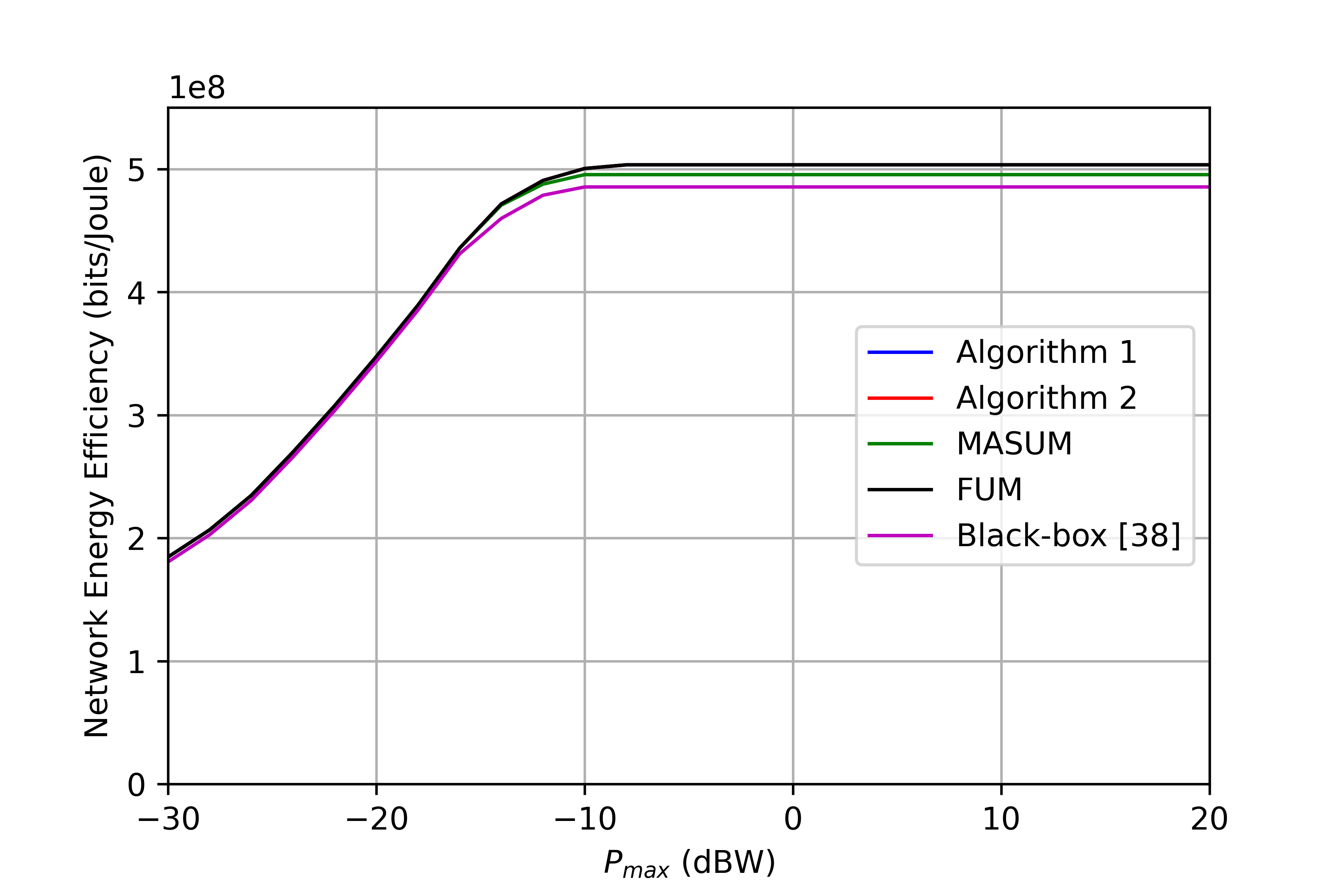}
\caption{Network energy efficiency for different values of $P_{max}$.}
\label{budget}
\end{figure}

Fig. \ref{budget} illustrates the energy efficiency performance for different values of $P_{max}$. We set the number of BSs to 7 and the number of users per BS is set to 4. The network energy efficiency is increasing with the increasing of $P_{max}$ and converges at $5.0342\times 10^8$ bits/Joule when $P_{max} = -9$ dBW. This means any increase in $P_{max}$ will lead only to increasing interference and the consumption and will not help in improving the rate. The performance of Algorithm 1, Algorithm 2, and FUM is identical while MASUM has slightly lower performance. Black-box model achieves the worst performance among all.

%

\subsection{Problem Size}
\begin{figure}[htp]
\centering
\includegraphics[width=3.5in]{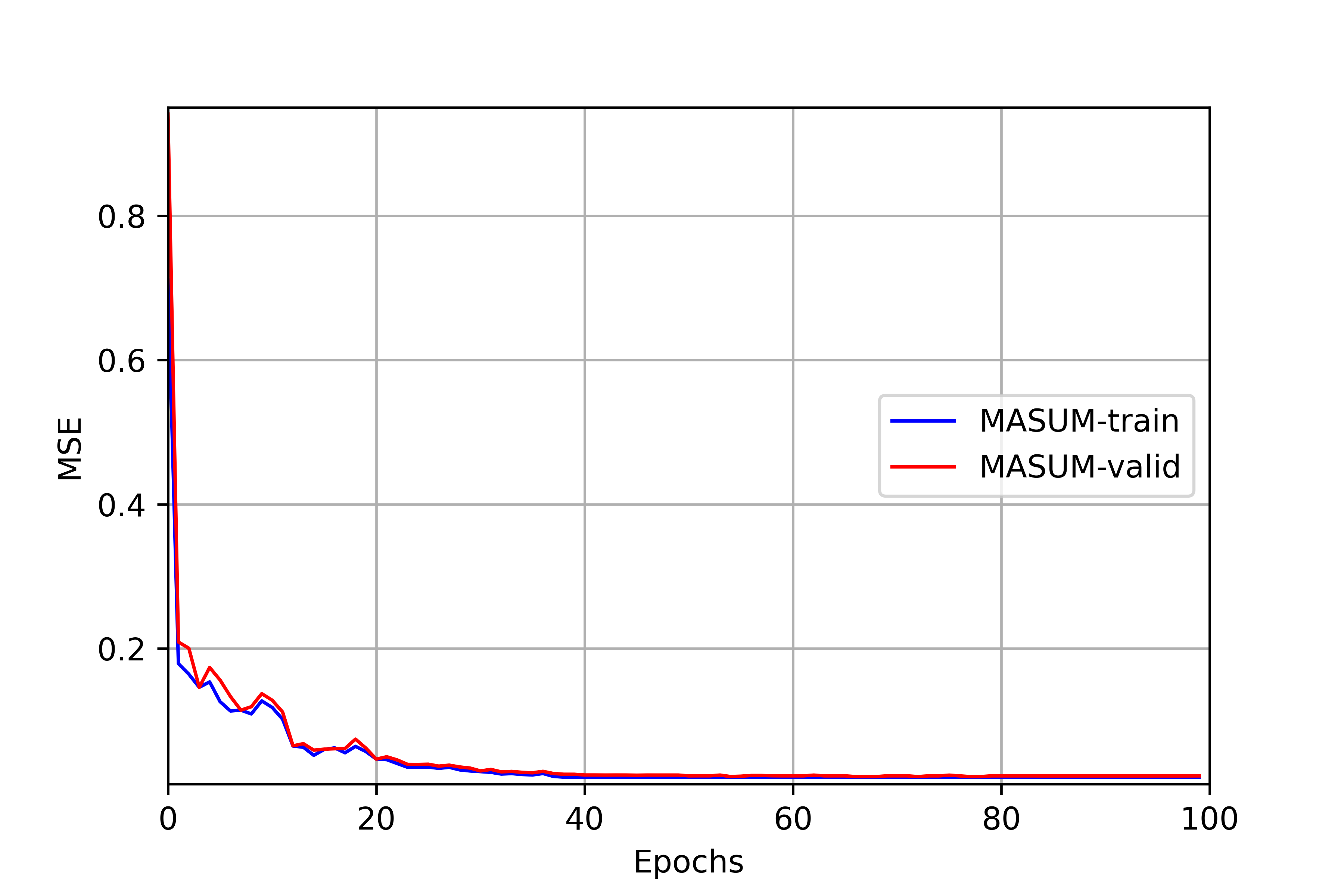}
\caption{Accuracy of MASUM for small problem (2 users per BS).}
\label{mseMASUM}
\end{figure}

For smaller problem when we have 7 BSs and each BS is serving 2 users, the performance improves dramatically. We reduced the size of the model to 6 layers in the main pipeline while including the attention in the last 5 layers and trained it for shorter training sequence. In Fig. \ref{mseMASUM}, we can notice that the accuracy of MASUM is high to 99.01 \% and 98.94\% respectively for training and validation.

\begin{figure}[htp]
\centering
\includegraphics[width=3.5in]{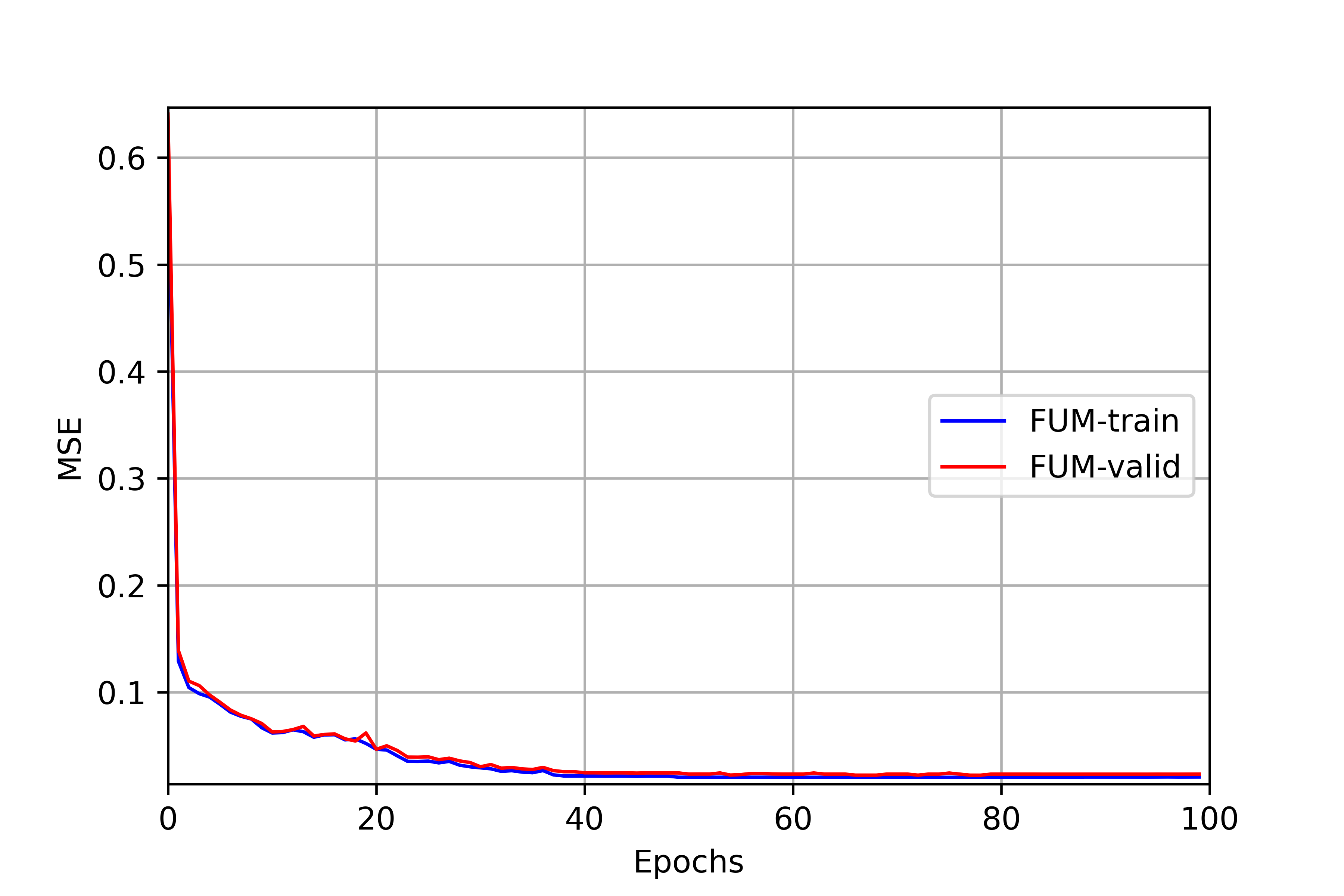}
\caption{Accuracy of FUM for 2 users per BS.}
\label{mseFUM}
\end{figure}

In case of FUM, we have 6 layers and we apply it on small problem similar to that in last case where we have 7 BSs; each BS is serving 2 users. From Fig. \ref{mseFUM} FUM achieves accuracy 99.37\% and 99.34\% respectively for training and validation.

\begin{figure}[htp]
\centering
\includegraphics[width=3.5in]{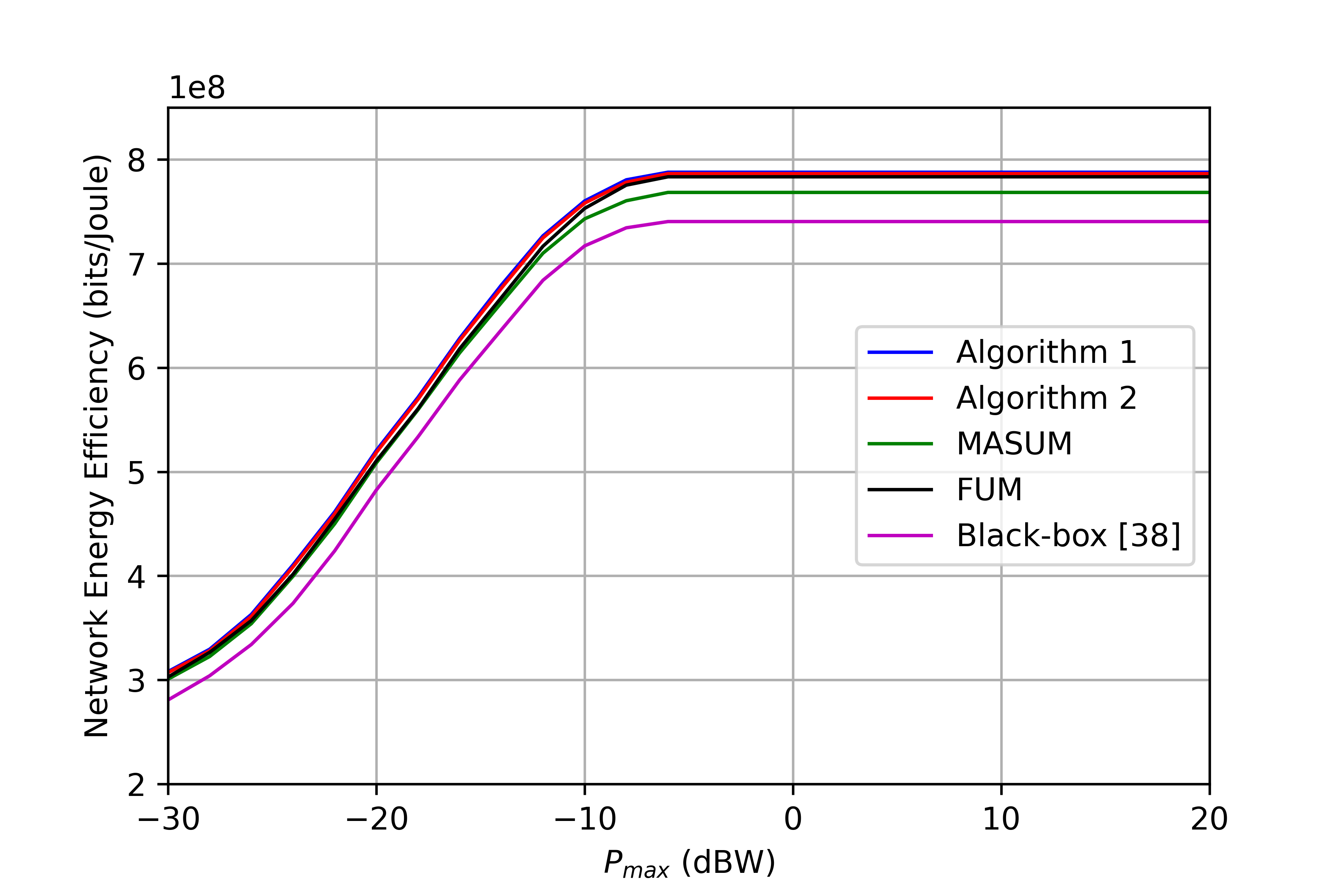}
\caption{Network energy efficiency for different values of $P_{max}$ 7 BSs and 8 users per BS.}
\label{bigbudget}
\end{figure}

Compared with the procedure in Fig. \ref{budget}, we test the models over large problem as in Fig. \ref{bigbudget}. We set the number of BS to 7 while the number of users per BS is set to 8. Considering the case of MASUM and FUM, we can observe that there is a degradation in the performance of both models. However, unlike the case of MASUM, this degradation is smaller in case of FUM. In case of MASUM, this degradation might be alleviated by increasing the number of neurons, the layers, and the attention blocks. Similarly, the increase of the number of neurons can decrease the impact of this degradation. It is worth mentioning that the adopted training procedure in this study has great impact on solidifying the performance of both models. MASUM achieves 2.4\% performance less than Algorithm 1 while FUM achieves about 1\% performance less than Algorithm 1. Although Black-box has the capability of learn the statistical information, its performance deteriorates with larger problems as in Fig. \ref{bigbudget}. Black-box achieves about 6.8\% performance less Algorithm 1.

\begingroup
\setlength{\tabcolsep}{10pt}
\renewcommand{\arraystretch}{1.5}
\begin{table*}[ht]
\centering
\caption{ABLATION STUDY ON THE IMPACT OF MODEL'S NUMBER OF STAGES/LAYERS}
\label{table2}
\begin{tabular}{|c|c|c|c|c|}
\hline
  \multirow{2}{*}{Model} & \multirow{2}{*}{No. of stages} & \multirow{2}{*}{Prediction accuracy (\%)} & \multicolumn{2}{c|}{Inference speed (ms)}\\\cline{4-5}
  &&&CPU&GPU\\ 
\hline
\multirow{3}{*}{MASUM}&5&98.55&0.0692&0.0571\\\cline{2-5}
&9&98.80&0.0823&0.0615\\\cline{2-5}
&14&98.80&0.0936&0.0652\\
\hline
\multirow{3}{*}{FUM}&5&99.11&0.0513&0.0420\\\cline{2-5}
&9&99.32&0.0534&0.0464\\\cline{2-5}
&14&99.32&0.0545&0.0474\\
\hline
\end{tabular}
\end{table*}
\endgroup


\subsection{Inference over Off-training Input}
\begin{figure}[htp]
\centering
\includegraphics[width=3.5in]{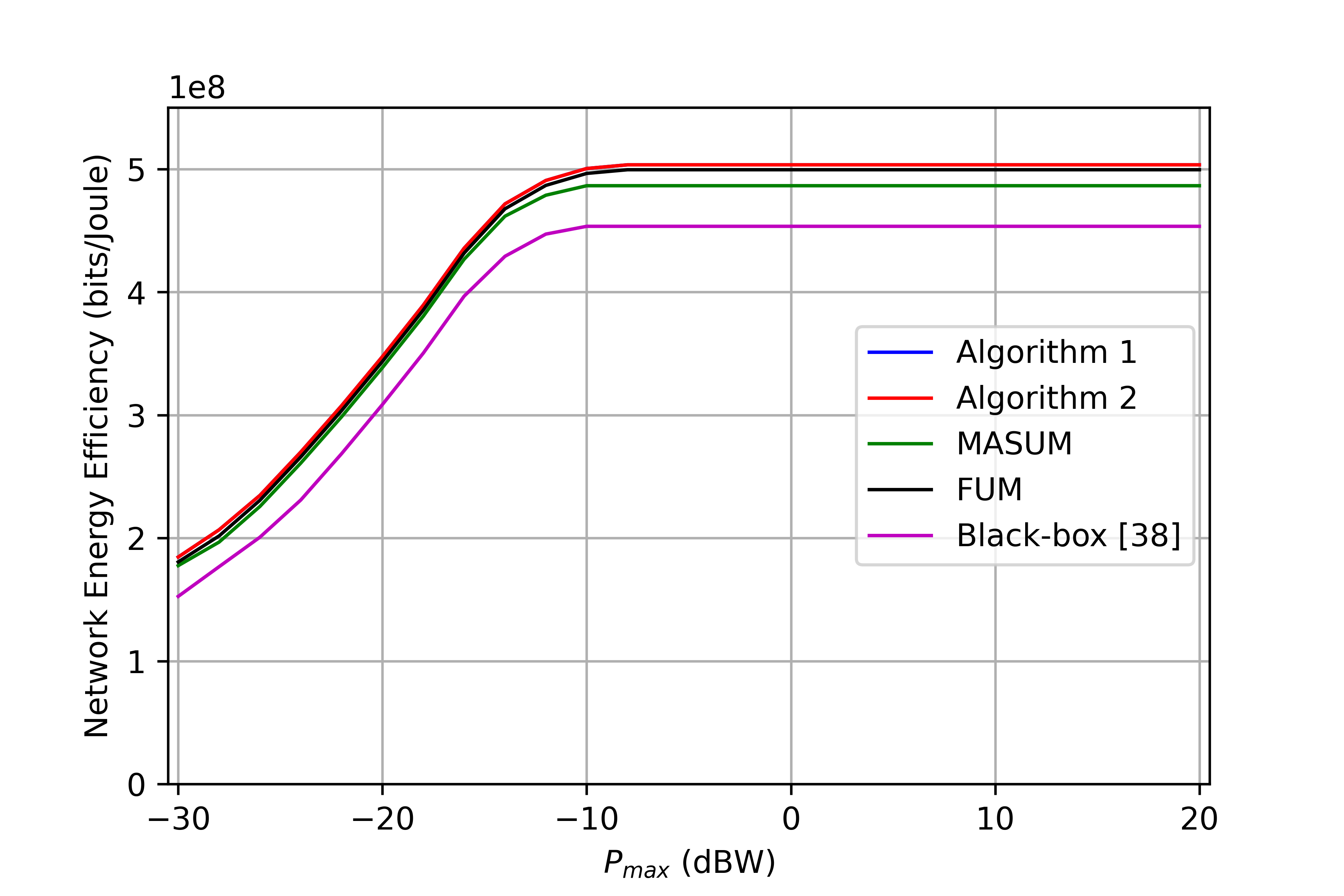}
\caption{Performance of different models over off-training data.}
\label{offtraining}
\end{figure}

Taking into consideration the changes in the channel conditions in the real-world, we test the schemes with off-training data and the result is shown as in Fig. \ref{offtraining}. We set the number of BSs to 7 and the number of users per BS is set to 4. Although there is a deterioration in the performance of FUM and MASUM compared with Algorithm 1 and Algorithm 2, the proposed two deep learning models show robustness against the changes in the channel conditions. FUM employs the full domain knowledge of the problem and learns the statistical information from the data. Therefore, FUM performs better than MASUM. FUM achieves 99.20\% performance of Algorithm 1 while MASUM achieves 96.57\% performance of Algorithm 1. Finally, Black-box performance significantly degrades and only 89.55\% of Algorithm 1 can be achieved.

\subsection{Ablation Study}

According to the deep unfolding concept, the number of the layers of the resulting model is mostly expected to be equal to the number of the iterations of the traditional algorithm. However, the number of the required layers to achieve the maximum possible accuracy usually does not depend on the number of iterations. Hence, we implement an ablation study to investigate the optimal structure to build each of the proposed MASUM and FUM models. TABLE \ref{table2} investigates the optimal number of layers to obtain the maximum possible accuracy while maintaining the inference speed at reasonable level. In case of FUM, we can observe that five to six layers are enough to obtain excellent performance. For instance, for five layers, the accuracy is 99.11\% and the model delivers the output in 0.0513 ms and 0.0420 ms respectively in CPU and GPU. However, the larger number of layers, take 9 layers, we obtain accuracy of 99.32\% with inference speed 0.0534 ms and 0.0464 ms for CPU and GPU, respectively. Obviously, increasing the number of layers beyond 6 layers will only lead to increase the inference speed with any improvement in the inference speed. It is also worth mentioning that the adopted training mechanism has great impact in tuning the model and optimizing the number of layers. Similar observations can be drawn in the case MASUM. However, since MASUM is of a hybrid nature in which model-driven and data-driven mechanisms are fused, other factors such as structure of the data-driven blocks and the position of these blocks have crucial impact on the performance.

\begingroup
\setlength{\tabcolsep}{10pt}
\renewcommand{\arraystretch}{1.5}
\begin{table*}[ht]
\centering
\caption{IMPACT OF THE NUMBER OF ATTENTION BLOCKS}
\label{table3}
\begin{tabular}{|c|c|c|c|}
  \hline
  No. of attention blocks & Accuracy (\%) & Inference speed (ms) & Location of attention blocks \\
  \hline
  0 & 97.92 & 0.0052 & - \\
  \hline
  2 & 98.68 & 0.0588 & 4, 5 \\
  \hline
  3 & 98.55 & 0.0602 & 3,4,5 \\
  \hline
  5 & 98.80 & 0.0615 & 5,6,7,8,9 \\
  \hline
  7 & 98.80 & 0.0711 & 8,9,10,11,12,13,14 \\
  \hline
\end{tabular}
\end{table*}
\endgroup
TABLE \ref{table3} illustrates the impact of the attention block on the performance of MASUM. First, it is obvious that the number of the usable attention block is dependent on the number of the overall layers. Second, the locations of the attention blocks have impact on the accuracy because it employ the changes in the features and create spatial connection and better context for prediction. Therefore, in the case of have five-layer MASUM, we deploy the attention blocks in the last two and three layers (depending on the number of the deployed attention blocks). From TABLE \ref{table3}, we can observe that the reasonable number of attention blocks is 5 if we considered the 7-layer structure. Also it is obvious that the inference speed decreases with the increase of number of attention blocks.

\section{Conclusions and Future Works}
In this work, we designed and compared two deep unfolding-based models for power control in multiple interference links. The problem was formulated as SoRP which is nonconvex. We considered two solution scenarios for the solution. First, we employed multidimensional fractional programming to obtain a numerical solution. Second, we combined multidimensional fractional programming and Lagrange dual transform to obtain a closed-form solution for the problem. On the highlights of the two solutions, we designed two deep unfolding-based models. The first model (namely MASUM) is based on the numerical solution where two pipelines were designed to predict the power. The first pipeline emulates the iterations of the iterative algorithm with the help of some data-driven layers. Due to the loss in the main pipeline, a second pipeline consists of attention subnets is incorporated to compensate the loss. Due to the presence the closed-form solution, the second model-driven neural network (namely FUM) can fully emulate the iterative solution with neither convolution nor attention subnet. Both MASUM and FUM achieved high accuracy in terms of power allocation with spectacular inference speed compared to the iterative solutions. We conclude that the presence of closed-form solution can facilitate unfolding the iterative solution into model-driven neural network. Furthermore, in case of difficulty of fully unfolding the iterative solution, we resort to semi-unfolding where we can partially unfold the iterative solution then employ the recent advances in data-driven models to enhance the performance. In future, we consider designing improved model-driven neural networks for large size problems using architectures such as inception-residual. To make the future model more suitable for dynamic environment and sparsity of wireless communication networks, we aim to incorporate the idea of liquid neural networks when building the model.

\section*{Appendix A}
\section*{Proof of Theorem 1}

Let us consider the case of Algorithm 1. Let $f\left( p \right) = \sum\limits_{m = 1}^M {\sum\limits_{k = 1}^K {{\omega _{m,k}}{\eta _{m,k}}} } $. We know from the definition of $\eta _{m,k}$ it is nondecreasing SoR. Hence, the problem \eqref{eqn6:main} can be conveniently considered equivalent to the following:
\begin{equation}
\begin{aligned}
& \underset{\rho,y}{\text{maximize}}
& & f\left( {p,y} \right)\\
& \text{subject to }
& &\eqref{eqn6:a}, \eqref{eqn8:a},\notag
\end{aligned}
\label{app1}
\end{equation}
where $f\left( {p,y} \right) = \sum\limits_{m = 1}^M {\sum\limits_{k = 1}^K {{\omega _{m,k}}\left( {2{y_{m,k}}\sqrt {{R_{m,k}}}  - y_{m,k}^2\left( {{p_{m,k}} + {p_{k,c}}} \right)} \right)} } $. Furthermore, we can rewrite the \eqref{eqn6:main} as a function of $\eta$ as below:
\begin{equation}
\begin{aligned}
& \underset{\rho,y}{\text{maximize}}
& & f\left( \eta \right)\\
& \text{subject to }
& &\eqref{eqn6:a},\notag\\
&&&{\eta _{m,k}} = \frac{{{R_{m,k}}}}{{{\rho_{m,k}}P_{max} + {p_{k,c}}}},
\end{aligned}
\label{app2}
\end{equation}

Hence, according to the quadratic transform theory in \cite{bb53}, we can replace the variable $\eta$  with ${\omega _{m,k}}\left( {2{y_{m,k}}\sqrt {{R_{m,k}}}  - y_{m,k}^2\left( {{\rho_{m,k}}P_{max} + {p_{k,c}}} \right)} \right)$ and since $f\left( \eta \right)$ is nondecreasing, $\mathop {\max }\limits_p \sum\limits_{m = 1}^M {\sum\limits_{k = 1}^K {\mathop {\max }\limits_y \left[ {{\omega _{m,k}}\left( {2{y_{m,k}}\sqrt {{R_{m,k}}}  - y_{m,k}^2\left( {{\rho_{m,k}}P_{max} + {p_{k,c}}} \right)} \right)} \right]} } $ can be rewritten as \eqref{app1}.

Next, we recast the data rate term as concave as in \eqref{eqn10}. Consequently, we obtain problem \eqref{eqn12:main}. At each iteration, and when $z_{m,k}$ and $y_{m,k}$ are fixed, the solution represents a stationary point in which Algorithm 1 converges. Similar reasoning can be applied in case of Algorithm 2.

\ifCLASSOPTIONcaptionsoff
  \newpage
\fi

\end{document}